# Anomalous Conductance Oscillations in the Hybridization Gap of InAs/GaSb Quantum Wells

Zhongdong Han[1], Tingxin Li[2], Long Zhang[3,*], Gerard Sullivan[5] and Rui-Rui Du[1,2,4,†]

[1]*International Center for Quantum Materials, School of Physics, Peking University, Beijing 100871, China*
[2]*Department of Physics and Astronomy, Rice University, Houston, Texas 77251-1892, USA*
[3]*Kavli Institute for Theoretical Sciences and CAS Center for Excellence, University of Chinese Academy of Sciences, Beijing 100190, China*
[4]*Collaborative Innovation Center of Quantum Matter, Beijing 100871, China*
[5]*Teledyne Scientific and Imaging, Thousand Oaks, California 91603, USA*

[*]Long Zhang: longzhang@ucas.ac.cn; [†]Rui-Rui Du: rrd@rice.edu

## Abstract

We observe the magnetic oscillation of electric conductance in the two-dimensional InAs/GaSb quantum spin Hall insulator. Its insulating bulk origin is unambiguously demonstrated by the antiphase oscillations of the conductance and the resistance. Characteristically, the in-gap oscillation frequency is higher than the Shubnikov-de Haas oscillation close to the conduction band edge in the metallic regime. The temperature dependence shows both thermal activation and smearing effects, which cannot be described by the Lifshitz-Kosevich theory. A two-band Bernevig-Hughes-Zhang model with a large quasiparticle self-energy in the insulating regime is proposed to capture the main properties of the in-gap oscillations.

***Introduction.-*** Magnetic oscillations in metals stem from the Landau quantization of charged particles in magnetic field, and have been a standard tool to measure the Fermi surfaces of metals [1]. In two-dimensional electron systems (2DES), the Shubnikov-de Haas (SdH) oscillation of conductance evolves into the integer quantum Hall effect when only a few Landau levels (LLs) are filled. In the fractional quantum Hall effect, the magnetic oscillations of composite fermions offer an unique window looking into many-body physics in this strongly interacting electron system [2,3].

Recently, unconventional magnetization and resistivity oscillations were observed in the Kondo insulators, $SmB_6$ and $YbB_{12}$ [4-8], which challenged the canonical theory of magnetic oscillations and triggered intense studies and controversies. The first concern is whether these oscillations come from the insulating bulk states. This is obscured by the presence of metallic surface states in the 3D topological Kondo insulators [4,9,10]. Therefore, it is particularly desirable to observe magnetic oscillations in 2D topological insulators, e.g., the InAs/GaSb quantum well (QW) [11], in which the bulk and edge transport channels can be clearly distinguished by designing different sample geometries. This is achieved in this work, and the insulating bulk state is shown to be the origin of the in-gap conductance oscillations.

Second, it is not clear whether the in-gap oscillations can be captured by the Landau quantization of gapped single-particle states [12-15] or whether one must consider some kind of (nearly) gapless charge-neutral excitations [16-20]. In experiments, the absence of low-temperature thermal conductivity [21,22] (cf. Refs. [6,23]) poses a severe constraint on charge-neutral excitations. On the other hand, in single-particle models,

• 2

the in-gap oscillations require a narrow hybridization gap [12-14]. The electron interactions and disorders can introduce a finite quasiparticle self-energy and reduce the hybridization gap, thereby significantly enhance the in-gap oscillations [15,23].

In this work, we report the observation of magnetic oscillations in the 2D InAs/GaSb quantum spin Hall (QSH) insulator. The magnetic oscillations are observed both in the resistance of a Hall bar and in the conductance of a Corbino disk device. They exhibit a $\pi$ phase difference. This demonstrates the insulating bulk origin of the in-gap oscillations. The in-gap oscillation frequency is higher than the SdH oscillation in the conduction band close to the band edge. This is captured by a two-band Bernevig-Hughes-Zhang model with a large quasiparticle self-energy introduced phenomenologically in the insulating regime, which may be attributed to the unscreened Coulomb interaction. The temperature dependence of the in-gap oscillation strongly deviates from the conventional Lifshitz-Kosevich (LK) theory in metals [1,24]. The oscillation amplitude can either increase (thermal activation) or decrease (thermal smearing) with the temperature depending on the Fermi energy, which is qualitatively consistent with the theoretical predictions in narrow-gap topological insulators [11,13].

*Materials and methods.-* The InAs/GaSb QW is a type-II semiconductor heterostructure. The valence band edge of the GaSb layer is higher than the conduction band edge of the InAs layer. The hybridization of these spatially separated electron and hole bands opens a mini-gap [25]. It has been established to be a highly tunable platform to realize the 2D QSH effect [26-28].

The InAs/GaSb QWs are grown on the n-type GaSb substrate by molecular beam epitaxy. The detailed structure of the QW is shown in the Supplemental Materials. Both dual-gated Hall bars and Corbino disk devices are fabricated. Except for the $R_{xx}$ trace in Fig. 2a, all conductance oscillations are measured from a Corbino device. The measurements are performed using the standard low-frequency lock-in technique in a $^3$He cryostat equipped with a superconducting magnet coil up to 9 T. We also measure the conductance in tilted magnetic field in a $^3$He cryostat equipped with a superconducting magnet up to 15 T. All data are taken at 300 mK except for the temperature dependence measurements.

We also fabricate the strained-layer InAs/Ga$_{0.68}$In$_{0.32}$Sb QSH insulator, which has a larger hybridization gap and a more insulating bulk state [29,30]. Similar in-gap resistance oscillations are observed and described in detail in the Supplemental Materials.

***In-gap magnetic oscillations from insulating bulk state.-*** We first show a standard Landau fan diagram to visualize the LLs of the InAs/GaSb QW in a perpendicular magnetic field $B_\perp$. A 2D map of the conductance $G_{xx}$ measured with a Corbino device as a function of the front gate bias $V_f$ and $B_\perp$ is shown in Fig. 1a. The gating $V_f$ tunes the Fermi energy from the conduction band to the valence band across the hybridization gap. One representative $G_{xx}$ curve in each regime is drawn in Fig. 1c. The blue and the red curves exhibit typical SdH oscillations of 2D metals. Surprisingly, the conductance oscillation is also observed in the QSH insulator regime (black curve). While the Landau level spectra in inverted InAs/GaSb system have been reported in a number of interesting experimental papers [31-33], the present work will focus on the in-gap magnetic oscillations.

Does the in-gap conductance oscillation come from the insulating bulk state? This is a major issue not fully settled in 3D Kondo insulators due to the presence of metallic surface states [4,9,10]. Fortunately, the electric transport of the bulk and the edge states can be distinguished in 2D devices of the InAs/GaSb QWs. The conductance $G_{xx}$ and the resistance $R_{xx}$ are measured with a Corbino disk and a Hall bar device, respectively. They show antiphase oscillations with $B_\perp$ (Fig. 2a).

First, in the Corbino device, the edge states are shunted by the metal electrodes, thus the conductance is only contributed by the bulk state. Second, the bulk resistance reaches 200 kΩ per square at zero magnetic field, which is order-of-magnitude larger than the resistance quantum $h/e^2 = 25.8$ kΩ, and thus the bulk state is sufficiently insulating. In 2D metals, the Hall resistivity $\rho_{xy}$ is much larger than the longitudinal resistivity $\rho_{xx}$ in a moderate $B_\perp$, thus the conductivity $\sigma_{xx} = \rho_{xx}/(\rho_{xx}^2 + \rho_{xy}^2) \simeq \rho_{xx}/\rho_{xy}^2 \propto \rho_{xx}$ should show in-phase oscillations as $\rho_{xx}$. However, in an insulator, $\rho_{xy} \ll \rho_{xx}$, thus $\sigma_{xx} \simeq 1/\rho_{xx}$, leading to the antiphase oscillations of $G_{xx}$ and $R_{xx}$. Therefore, the above observations unambiguously demonstrate that the in-gap magnetic oscillations originate from the insulating bulk state.

The in-gap conductance oscillation is approximately periodic in $1/B$ like the SdH effect in metals. In single-particle models of magnetic oscillations in hybridized insulators, the oscillation frequency is determined by the electron density $n_0$ of the compensated semimetal if the band hybridization is turned off [12-15,23],

$$\frac{\Delta \nu}{\Delta(1/B)} = \frac{h}{e} n_0, \qquad (1)$$

in which $\nu$ is the filling factor. We plot the filling factor diagram of the conductance minima in Fig. 2b. The fitting yields $n_0 = (1.63 \pm 0.05) \times 10^{11}$ cm$^{-2}$. Quite unexpectedly, the electron density $n$ exhibits a jump when $V_f$ is tuned from the conduction band into the hybridization gap (Fig. 2b, inset). In other words, the electron density in the conduction band extracted from the SdH oscillations can be smaller than $n_0$.

The InAs/GaSb QSH insulator is described by the two-band Bernevig-Hughes-Zhang (BHZ) model [26,34],

$$H_0 = \begin{pmatrix} h(k) & 0 \\ 0 & h^*(-k) \end{pmatrix},$$

$$h(k) = \begin{pmatrix} \dfrac{\hbar^2 k^2}{2m_e^*} - \mu_e & w(k_x + ik_y) \\ w(k_x - ik_y) & -\dfrac{\hbar^2 k^2}{2m_h^*} + \mu_h \end{pmatrix}.$$

Here, the electron and hole effective masses $m_e^* = 0.040 m_e$ and $m_h^* = 0.136 m_e$ are taken in accord with the experiments [35]. The band inversion $\mu_e + \mu_h = \pi \hbar^2 n_0 (1/m_e^* + 1/m_h^*)$. We set $w = 0.8$ eV·Å, which is comparable to the results of first-principle calculations with similar QW widths [36].

The band structure of the BHZ model is shown in Fig. 2c. When the Fermi energy is in the conduction band, the SdH oscillation frequency is proportional to the electron density $n_e$, $\dfrac{\Delta \nu}{\Delta(1/B)} = \dfrac{h}{e} n_e$, which vanishes as the Fermi energy decreases to the band bottom.

In 2DES, the Coulomb interactions lead to a finite quasiparticle self-energy $\Sigma$, which should be quite large in the insulating regime due to the absence of charge screening. Here we treat $\Sigma = \text{diag}(-i\Gamma_e, -i\Gamma_h, -i\Gamma_e, -i\Gamma_h)$ as phenomenological

· 6

parameters, and consider the asymmetric case $\Gamma_h > \Gamma_e$ because of the larger hole mass. As shown by H. Shen and L. Fu [15], the asymmetric self-energy reduces the hybridization gap (see Fig. 2c) and greatly enhances the in-gap oscillation, and the frequency is given by Eq. (1). Therefore, the oscillation frequency should exhibit a jump compared to the SdH frequency in the metallic regime.

The density of states (DOS) at the Fermi energy in magnetic field are calculated in both the metallic and the insulating regimes assuming different self-energies, $D(\epsilon) = 2\text{Im}\text{Tr}(H_0 + \Sigma - \epsilon)^{-1}$, and the details are shown in the Supplemental Materials. The calculated carrier density $n$ is found to exhibit the similar variation as the experiments (Fig. 2d). We note that such a variation cannot be produced if the self-energy is taken to be the same in both regimes (see the Supplemental Materials). Therefore, this peculiar variation of the frequency points to the strong Coulomb interaction effect particularly in the insulating regime.

The band inversion of the QW can be tuned by the back gate $V_b$. The evolution of the in-gap conductance is shown in Fig. 4a. As $V_b$ is tuned from $3\text{ V}$ to $-1\text{ V}$, the band inversion gradually decreases and all oscillatory features shift to lower $B_\perp$, thus the oscillation frequency is reduced. This is consistent with Eq. (1) from the two-band model.

*Temperature dependence.-* The temperature dependence of the in-gap conductance is shown in Fig. 3. In order to extract the oscillation amplitudes, we take the second-order derivative $-\partial^2_{B^{-1}} G_{xx}$ to remove the monotonic background, which is shown in Figs. 3b and 3c. As the temperature increases, the in-gap oscillation amplitude near the valence band edge ($V_f = -0.85$ V) gradually decreases (Fig. 3b), which is reminiscent of the LK theory of thermal smearing effect in metals [1,24]. However, at the charge neutral point (CNP) ($V_f = -0.75$ V), the oscillation amplitude increases with the temperature (Fig. 3c), suggesting a thermal activation behavior, which cannot be described by the LK theory. Therefore, the temperature effect on the in-gap oscillations is two-fold, thermal smearing and activation, which is qualitatively consistent with the theory of DOS oscillations of in narrow-gap insulators [11,13].

***Robustness to in-plane magnetic field.-*** Applying an in-plane magnetic field $B_\parallel$ on type-II semiconductor QWs is supposed to induce a relative momentum shift between the electron and the hole bands, $\Delta k = eB_\parallel \langle z \rangle / \hbar$, in which $\langle z \rangle$ is the vertical distance of the electron and the hole layers. This may close the hybridization gap [25]. However, we find that the in-plane magnetic field (up to 12T) has weak effect on the in-gap oscillations in the sample studied (Fig. 4b).

The robustness against $B_\parallel$ may be understood by considering the effect of the interband Coulomb interaction. It can reduce the spatial separation $\langle z \rangle$ of electrons and holes and the momentum shift $\Delta k$, thus making the hybridization gap more robust against $B_\parallel$. Moreover, it can bind the electrons and holes into excitons, leading to a topological excitonic insulator (TEI) state [37,38]. In experiments, the TEI state is found in InAs/GaSb QWs with $n_0 < 1 \times 10^{11}$ cm$^{-2}$ [39], and the exciton gap is remarkably robust against $B_\parallel$. However, in Ref. [39], $n_0$ was estimated by linearly extrapolating $n$ in the conduction band to the CNP, which may underestimate $n_0$ according to our current analysis. In our devices, $n_0 = (1.63 \pm 0.05) \times 10^{11}$ cm$^{-2}$, thus the TEI state may also set in and reinforce the robustness against $B_\parallel$.

***Summary and discussion.-*** In summary, the magnetic oscillation of conductance is observed in the hybridization gap of inverted InAs/GaSb QWs. The antiphase oscillations of the conductance and the resistance demonstrate their insulating bulk origin. The variation of the oscillation frequency with the front gate bias $V_f$ can be captured by a two-band BHZ model with the phenomenological assumption of a large quasiparticle self-energy in the insulating regime, which may be attributed to the strong Coulomb interaction. The temperature dependence of the in-gap oscillation amplitudes shows both thermal smearing and thermal activation behaviors, which strongly deviates from the conventional LK theory. The in-plane magnetic field has rather weak effect on the in-gap oscillations.

In a recent preprint by D. Xiao et al [40], the in-gap resistance oscillation in wider InAs/GaSb QWs was also reported; the QWs have the width 15 nm of InAs layer and 10 nm of GaSb layer. An oscillation frequency jump was also found when the gate voltage tunes the Fermi energy from the conduction band to the hybridization gap. Their measurements were only performed on Hall-bar devices, and the electron density at the CNP is $n_0 \simeq 2 \times 10^{12}$ cm$^{-2}$, much larger than our sample.

## Acknowledgements

We gratefully acknowledge C. L. Yang and P. L. Li for the assistance in titled magnetic field experiments. This work is supported by National Key R&D Program of China (No. 2017YFA0303301 and No. 2018YFA0305800), Strategic Priority Research Program of Chinese Academy of Sciences (No. XDB28000000), National Natural Science

Foundation of China (No. 11804337), and Beijing Municipal Science & Technology Commission (No. Z181100004218001). Work at Rice University is supported by NSF Grant No. DMR-1508644 and Welch Foundation Grant No. C-1682.## References

[1] D. Shoenberg, *Magnetic Oscillations in Metals* (Cambridge University Press, Cambridge, 1984).

[2] B. I. Halperin, P. A. Lee, and N. Read, Phys. Rev. B 47, 7312(1993).

[3] J. K. Jain, *Composite Fermions* (Cambridge University Press, Cambridge, 2007).

[4] G. Li, Z. Xiang, F. Yu, T. Asaba, B. Lawson, P. Cai, C. Tinsman, A. Berkley, S. Wolgast, Y. S. Eo, D.-J. Kim, C. Kurdak, J. W. Allen, K. Sun, X. H. Chen, Y. Y. Wang, Z. Fisk, and L. Li, Science 346, 1208 (2014).

[5] B. S. Tan, Y.-T. Hsu, B. Zeng, M. C. Hatnean, N. Harrison, Z. Zhu, M. Hartstein, M. Kiourlappou, A. Srivastava, M. D. Johannes, T. P. Murphy, J.-H. Park, L. Balicas, G. G. Lonzarich, G. Balakrishnan, and S. E. Sebastian, Science 349, 287 (2015).

[6] M. Hartstein, W. H. Toews, Y.-T. Hsu, B. Zeng, X. Chen, M. C. Hatnean, Q. R. Zhang, S. Nakamura, A. S. Padgett, G. Rodway-Gant, J. Berk, M. K. Kingston, G. H. Zhang, M. K. Chan, S. Yamashita, T. Sakakibara, Y. Takano, J.-H. Park, L. Balicas, N. Harrison, N. Shitsevalova, G. Balakrishnan, G. G. Lonzarich, R. W. Hill, M. Sutherland, and S. E. Sebastian, Nat. Phys. 14, 166 (2018).

[7] H. Liu, M. Hartstein, G. J. Wallace, A. J. Davies, M. C. Hatnean, M. D. Johannes, N. Shitsevalova, G. Balakrishnan, and S. E. Sebastian, J. Phys. Condens. Matter 30, 16LT01 (2018).
· 10 ·

[8] Z. Xiang, Y. Kasahara, T. Asaba, B. Lawson, C. Tinsman, L. Chen, K. Sugimoto, S. Kawaguchi, Y. Sato, G. Li, S. Yao, Y. L. Chen, F. Iga, J. Singleton, Y. Matsuda, and L. Li, Science 362, 65 (2018).

[9] O. Erten, P. Ghaemi, and P. Coleman, Phys. Rev. Lett. 116, 046403 (2016).

[10] J. D. Denlinger, S. Jang, G. Li, L. Chen, B. J. Lawson, T. Asaba, C. Tinsman, F. Yu, K. Sun, J. W. Allen, C. Kurdak, D.-J. Kim, Z. Fisk, and L. Li, arXiv:1601.07408.

[11] J. Knolle and N. R. Cooper, Phys. Rev. Lett. 118, 176801 (2017).

[12] J. Knolle and N. R. Cooper, Phys. Rev. Lett. 115, 146401 (2015).

[13] L. Zhang, X.-Y. Song, and F. Wang, Phys. Rev. Lett. 116, 046404 (2016).

[14] H. K. Pal, F. Piéchon, J.-N. Fuchs, M. Goerbig, and G. Montambaux, Phys. Rev. B 94, 125140 (2016).

[15] H. Shen and L. Fu, Phys. Rev. Lett. 121, 026403 (2018).

[16] G. Baskaran, arXiv:1507.03477.

[17] J. Knolle and N. R. Cooper, Phys. Rev. Lett. 118, 096604 (2017).

[18] O. Erten, P.-Y. Chang, P. Coleman, and A. M. Tsvelik, Phys. Rev. Lett. 119, 057603 (2017).

[19] D. Chowdhury, I. Sodemann, and T. Senthil, Nat. Commun. 9, 1766 (2018).

[20] I. Sodemann, D. Chowdhury, and T. Senthil, Phys. Rev. B 97, 045152 (2018).

[21] Y. Xu, S. Cui, J. K. Dong, D. Zhao, T. Wu, X. H. Chen, K. Sun, H. Yao, and S. Y. Li, Phys. Rev. Lett. 116, 246403 (2016).

[22] M.-E. Boulanger, F. Laliberté, M. Dion, S. Badoux, N. Doiron-Leyraud, W. A. Phelan, S. M. Koohpayeh, W. T. Fuhrman, J. R. Chamorro, T. M. McQueen, X. F. Wang,

Y. Nakajima, T. Metz, J. Paglione, and L. Taillefer, Phys. Rev. B 97, 245141 (2018).

[23] N. Harrison, Phys. Rev. Lett. 121, 026602 (2018).

[24] I. M. Lifshitz and A. M. Kosevich, Sov. Phys. JETP 2, 636 (1956).

[25] M. J. Yang, C. H. Yang, B. R. Bennett, and B. V. Shanabrook, Phys. Rev. Lett. 78, 4613 (1997).

[26] C. Liu, T. L. Hughes, X.-L. Qi, K. Wang, and S.-C. Zhang, Phys. Rev. Lett. 100, 236601 (2008).

[27] I. Knez, R.-R. Du, and G. Sullivan, Phys. Rev. Lett. 107, 136603 (2011).

[28] L. Du, I. Knez, G. Sullivan, and R.-R. Du, Phys. Rev. Lett. 114, 096802 (2015).

[29] L. Du, T. Li, W. Lou, X. Wu, X. Liu, Z. Han, C. Zhang, G. Sullivan, A. Ikhlassi, K. Chang, and R.-R. Du, Phys. Rev. Lett. 119, 056803 (2017).

[30] T. Li, P. Wang, G. Sullivan, X. Lin, and R.-R. Du, Phys. Rev. B 96, 241406(R) (2017).

[31] R. J. Nicholas, K. Takashina, M. Lakrimi, B. Kardynal, S. Khym, N. J. Mason, D. M. Symons, D. K. Maude, and J. C. Portal, Phys. Rev. Lett. 85, 2364 (2000).

[32] K. Suzuki, K. Takashina, S. Miyashita, and Y. Hirayama, Phys. Rev. Lett. 93, 016803 (2004).

[33] Fabrizio Nichele, Atindra Nath Pal, Patrick Pietsch, Thomas Ihn, Klaus Ensslin, Christophe Charpentier, and Werner Wegscheider, Phys. Rev. Lett. 112, 036802 (2014).

[34] B. A. Bernevig, T. L. Hughes, and S.-C. Zhang, Science 314, 1757 (2006).

[35] X. Mu, G. Sullivan, and R.-R. Du, Appl. Phys. Lett. 108, 012101 (2016).

[36] C. Liu and S. Zhang, in *Topological Insulators*, edited by M. Franz and L.

Molenkamp (Elsevier, 2013).

[37] Y. Naveh and B. Laikhtman, Phys. Rev. Lett. 77, 900 (1996).

[38] D. I. Pikulin and T. Hyart, Phys. Rev. Lett. 112, 176403 (2014).

[39] L. Du, X. Li, W. Lou, G. Sullivan, K. Chang, J. Kono, and R.-R. Du, Nat. Commun. 8, 1971 (2017).

[40] D. Xiao, L.-H. Hu, C. Liu, and N. Samarth, arXiv:1812.05238.

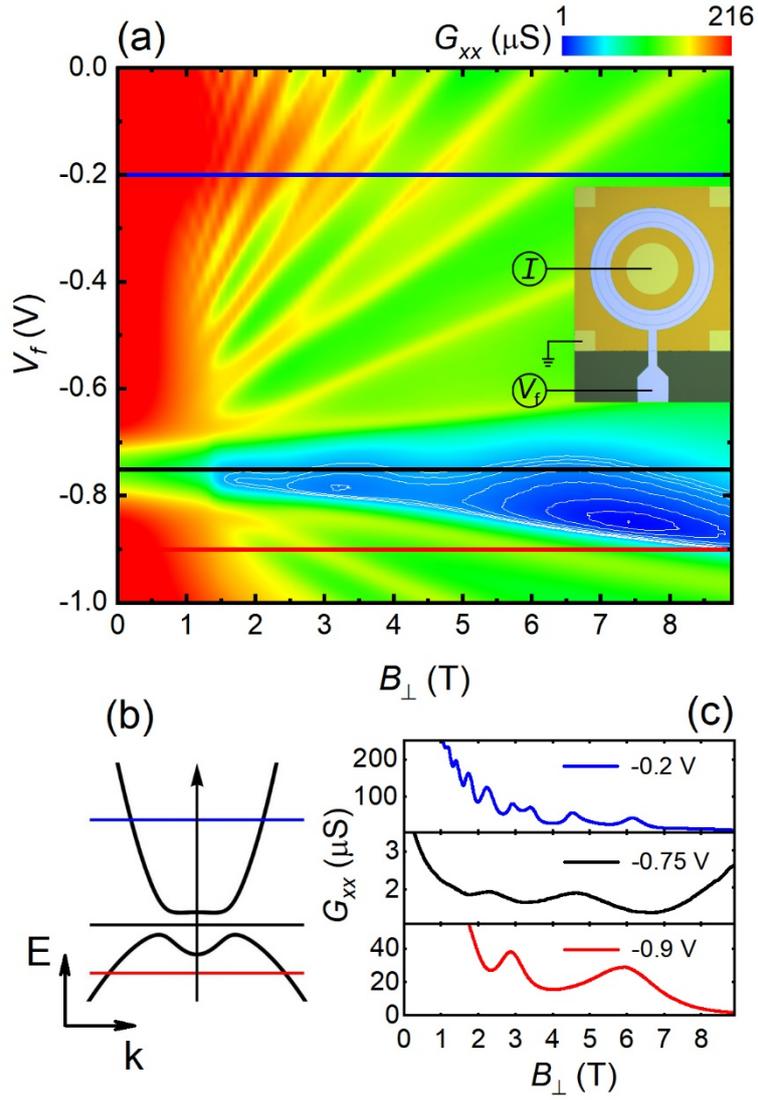

Fig. 1. (a) 2D conductance map as a function of the front gate bias $V_f$ and the perpendicular magnetic field $B_\perp$, which is measured with a Corbino device (optical microscope image in inset) at 300 mK. The conductance traces at three gate biases are plotted in (c) to represent the electron ($V_f = -0.2$ V, blue), the hole ($V_f = -0.9$ V, red) and the insulating ($V_f = -0.75$ V, black) regimes, respectively. The inverted band structure is schematically plotted in (b), where the Fermi energies of the three curves are labeled.

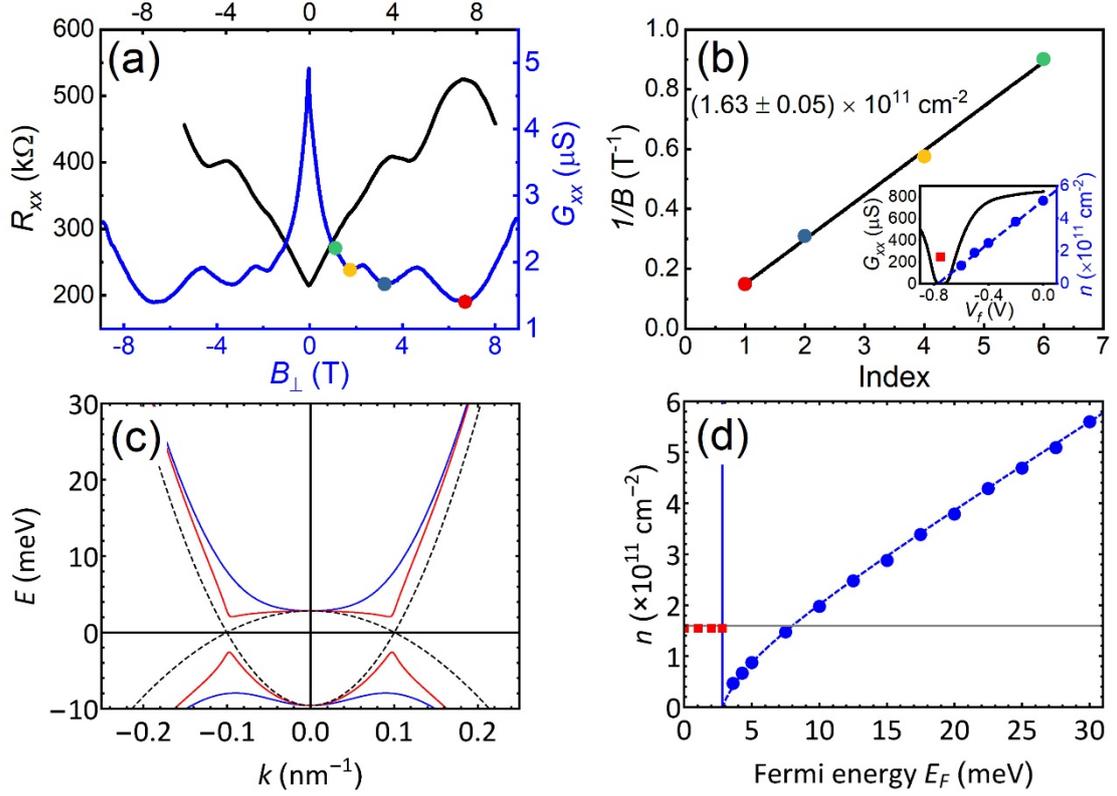

Fig. 2. (a) Anomalous magnetic oscillations of $G_{xx}$ and $R_{xx}$ in the hybridization gap. Two $B_\perp$-axes with different scales are used here to mitigate the mismatches due to the density difference from different devices. The top (bottom) axis is linked to left (right) axis. (b) Filling factor index diagram. The fitting yields $n_0 = (1.63 \pm 0.05) \times 10^{11}\ \text{cm}^{-2}$. Inset: Electron density $n$ in the conduction band (blue dots) and $n_0$ in the hybridization gap (red dot), and zero-field conductance $G_{xx}$ (black curve) vs. $V_f$. The dashed line is guide for the eye. (c) Energy dispersion (real part of the eigenvalues) of $H = H_0 + \Sigma$ with $n_0 = 1.6 \times 10^{11}\ \text{cm}^{-2}$. Blue lines: $\Gamma_e = \Gamma_h = 0$; red lines: $\Gamma_e = 1\ \text{meV}$, $\Gamma_h = 16\ \text{meV}$. Also shown is the dispersion without hybridization (black dashed lines). (d) Electron density fitted from the density of states oscillations calculated from the two-band model. Blue dots: in the metallic regime, $\Gamma_e = 0.2\ \text{meV}$, $\Gamma_h = 0.4\ \text{meV}$. Red squares: in the insulating regime, $\Gamma_e = 1\ \text{meV}$, $\Gamma_h = 16\ \text{meV}$. $n_0$ is indicated by the gray line. The blue vertical line indicates the conduction band edge. The dashed line is the Fermi surface area in the conduction band when the self-energy is turned off.

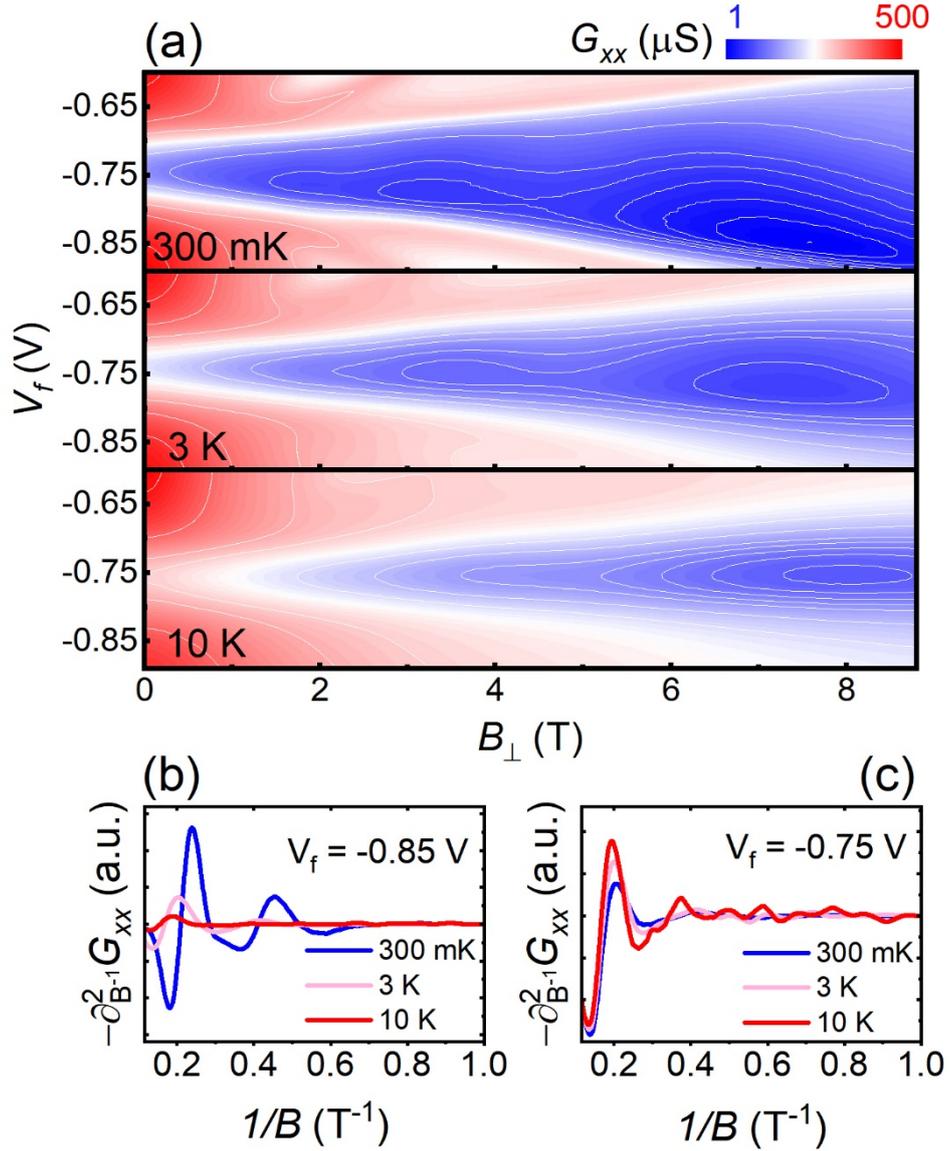

Fig. 3. Temperature dependence of the in-gap oscillations. (a) 2D conductance map in the insulator regime measured at different temperatures. In order to compare the oscillation amplitudes at different temperatures, we remove the background by taking the second-order derivative, $-\partial^2_{B^{-1}} G_{xx}$ in (b) at $V_f = -0.85$ V and (c) at $V_f = -0.75$ V.

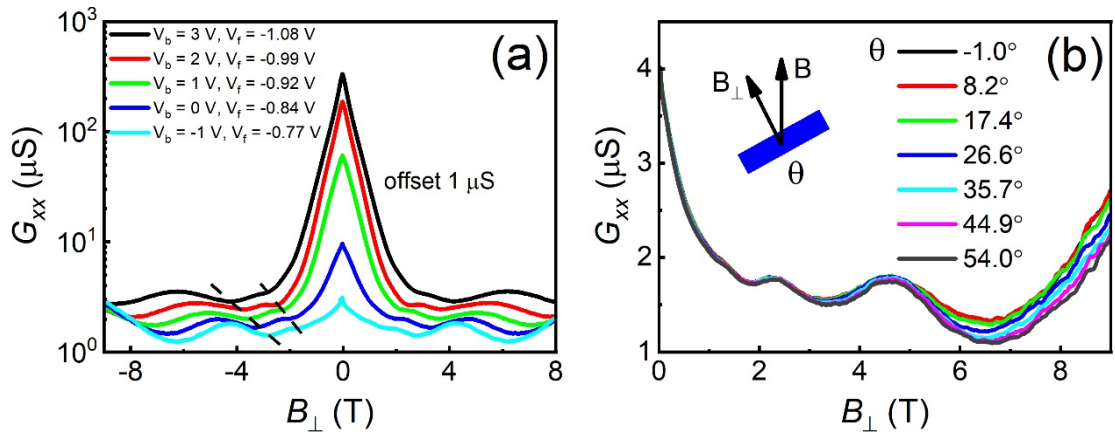

Fig. 4. (a) The in-gap oscillations measured with different back gate biases $V_b$. The band inversion gradually decreases as $V_b$ is tuned from $3\,\text{V}$ to $-1\,\text{V}$. (b) The in-gap oscillations measured in tilted magnetic fields.

# Supplemental Materials for *Anomalous Conductance Oscillations in the Hybridization Gap of InAs/GaSb Quantum Wells*

Zhongdong Han[1], Tingxin Li[2], Long Zhang[3,*], Gerard Sullivan[5] and Rui-Rui Du[1,2,4,†]

[1]*International Center for Quantum Materials, School of Physics, Peking University, Beijing 100871, China*
[2]*Department of Physics and Astronomy, Rice University, Houston, Texas 77251-1892, USA*
[3]*Kavli Institute for Theoretical Sciences and CAS Center for Excellence, University of Chinese Academy of Sciences, Beijing 100190, China*
[4]*Collaborative Innovation Center of Quantum Matter, Beijing 100871, China*
[5]*Teledyne Scientific and Imaging, Thousand Oaks, California 91603, USA*

*Long Zhang: longzhang@ucas.ac.cn; †Rui-Rui Du: rrd@rice.edu

**Wafer structure and device fabrication**

In our experiment, the samples used are grown by molecular beam epitaxy, with wafer structure shown in Fig. S1. QW of 11 nm wide InAs and 7 nm GaSb is sandwiched between two $Al_{0.7}Ga_{0.3}Sb$ barriers. An n-doped GaSb substrate is chosen to enhance the mobility of the sample and severed as a natural back gate. By processing with standard photo lithography and electron beam evaporation, a Corbino device with the radius of the inner and outer rings being 600 μm and 1200 μm is fabricated to measure the bulk conductance $G_{xx}$. There is no contribution of edge states, which are shunted in this geometry. Ohmic contacts are made with Ti/Au thin film and post annealing in the atmosphere of hydrogen-nitrogen forming gas. Following the deposition of $Al_2O_3$ layer on the entire surface, 100 nm Ti/Au is evaporated as a metallic front gate. A similar process applies to Hall bar devices as well.

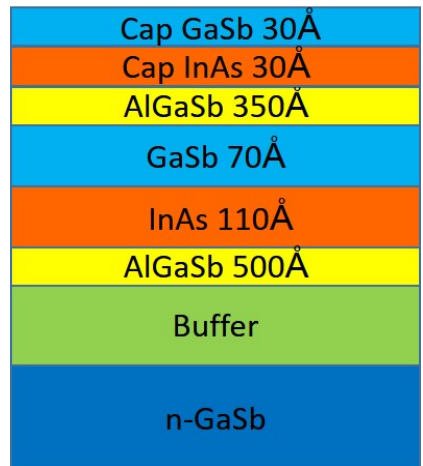

Fig. S1. Wafer structure of bilayer InAs/GaSb QWs

## In-gap resistance oscillations in strained-layer InAs/Ga$_{0.68}$In$_{0.32}$Sb QWs

We also observe similar in-gap oscillations in strained-layer InAs/GaInSb QWs. As an example, Fig. S2 (b) shows in-gap resistance oscillation in a 75 × 25 μm$^2$ Hall bar. Due to the in-plane strain effect, InAs/GaInSb system in general has a larger hybridization gap and a more insulating bulk state. The energy gap is Δ ~ 250 K in InAs/Ga$_{0.68}$In$_{0.32}$Sb QWs, which is about five fold of that in InAs/GaSb QWs [1]. According to current theoretical explanations, a narrow gap is necessary for observing anomalous quantum oscillations in an insulator [2-5]. The oscillation in a large gap insulator observed in our experiments may post a challenge to the current understanding and bring in new perspectives of the anomalous in-gap oscillations.

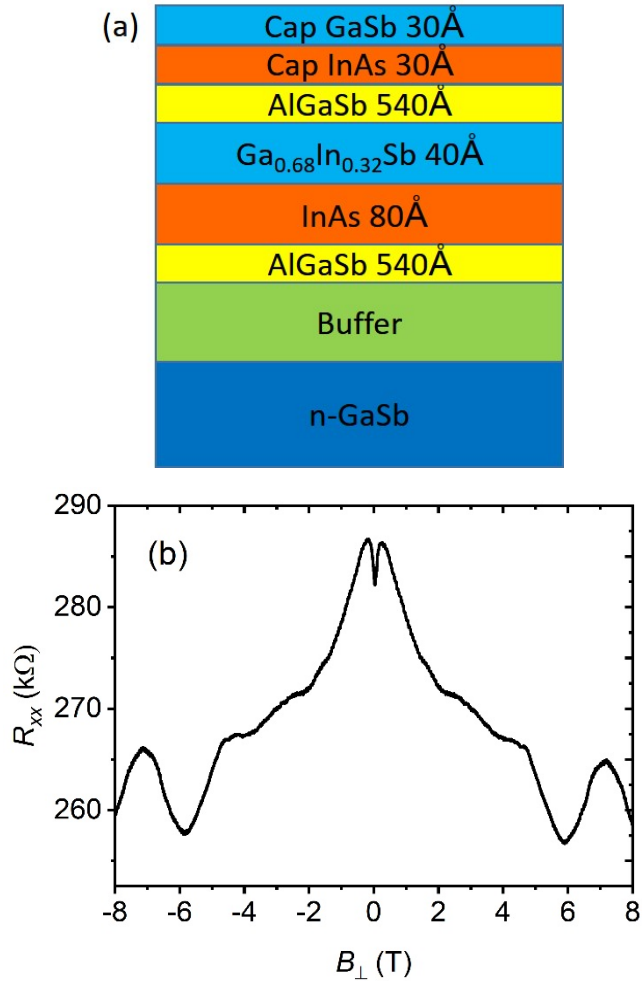

Fig. S2. (a) Wafer structure of the strained-layer InAs/Ga$_{0.68}$In$_{0.32}$Sb QWs. (b) In-gap resistance oscillation observed in a 75 × 25 μm$^2$ Hall bar made by InAs/Ga$_{0.68}$In$_{0.32}$Sb QWs.

## Theoretical calculations of density of states oscillations

The two-band Bernevig-Hughes-Zhang (BHZ) model for the InAs/GaSb quantum well is given in the basis of $\{|E+\rangle, |H+\rangle, |E-\rangle, |H-\rangle\}$ by [6,7]

$$H_0 = \begin{pmatrix} h(k) & 0 \\ 0 & h^*(-k) \end{pmatrix},$$

$$h(k) = \begin{pmatrix} \dfrac{\hbar^2 k^2}{2m_e^*} - \mu_e & w(k_x + ik_y) \\ w(k_x - ik_y) & -\dfrac{\hbar^2 k^2}{2m_h^*} + \mu_h \end{pmatrix}.$$

In a perpendicular magnetic field $B$, $\vec{k}$ is replace by $\vec{k} + \dfrac{e\vec{A}}{\hbar}$, and a Zeemann term should also be included,

$$H_Z = \text{diag}\left(-\dfrac{1}{2}g_e\mu_B B, -\dfrac{1}{2}g_h\mu_B B, \dfrac{1}{2}g_e\mu_B B, \dfrac{1}{2}g_h\mu_B B\right),$$

In which $g_e = 11.5$ and $g_h = 0.9$ are also taken in accord with the experiments [8].

In the $\{|E+\rangle, |H+\rangle\}$ sector, the $(n+1)$-th electron Landau level (LL) hybridizes with the $n$-th hole LL. The total Hamiltonian including the self-energy term is given by

$$h_+ = \begin{pmatrix} \hbar\omega_e\left(n + \dfrac{3}{2}\right) - \mu_e - \dfrac{1}{2}g_e\mu_B B - i\Gamma_e & w\sqrt{\dfrac{2eB(n+1)}{\hbar}} \\ w\sqrt{\dfrac{2eB(n+1)}{\hbar}} & -\hbar\omega_h\left(n + \dfrac{1}{2}\right) + \mu_h - \dfrac{1}{2}g_h\mu_B B - i\Gamma_h \end{pmatrix},$$

$n \geq 0$.

The zeroth electron LL is unhybridized with the energy eigenvalue

$$\epsilon_{0+} = \dfrac{1}{2}\hbar\omega_e - \mu_e - \dfrac{1}{2}g_e\mu_B B - i\Gamma_e.$$

In the $\{|E-\rangle, |H-\rangle\}$ sector, the $n$-th electron LL hybridizes with the $(n+1)$-th hole LL. The total Hamiltonian including the self-energy term is given by

$$h_- = \begin{pmatrix} \hbar\omega_e\left(n + \dfrac{1}{2}\right) - \mu_e + \dfrac{1}{2}g_e\mu_B B - i\Gamma_e & -w\sqrt{\dfrac{2eB(n+1)}{\hbar}} \\ -w\sqrt{\dfrac{2eB(n+1)}{\hbar}} & -\hbar\omega_h\left(n + \dfrac{3}{2}\right) + \mu_h + \dfrac{1}{2}g_h\mu_B B - i\Gamma_h \end{pmatrix},$$

$n \geq 0$.

The zeroth hole LL is unhybridized with the energy eigenvalue

$$\epsilon_{0-} = -\dfrac{1}{2}\hbar\omega_h + \mu_h + \dfrac{1}{2}g_e\mu_B B - i\Gamma_h.$$

The self-energy $\Gamma_e$ and $\Gamma_h$ are treated as phenomenological parameters as discussed in the main text. The degeneracy of each LL is given by the number of magnetic flux quanta penetrating the two-dimensional electron system. The density of states (DOS) at a fixed Fermi energy $\epsilon_F$ is calculated by $D(\epsilon_F) = 2\mathrm{Im}\mathrm{Tr}(H_0 + H_Z + \Sigma - \epsilon_F)^{-1}$.

## A. DOS oscillations in the metallic regime

We first calculate the DOS in magnetic fields without including the Zeemann term. The self-energy is set to be $\Gamma_e = 0.2$ meV, and $\Gamma_h = 0.4$ meV. The DOS oscillation is clearly exhibited in Fig. S3, and the oscillation frequency is shown in Fig. 2d in the main text. We have also verified that the Zeemann term does not change the oscillation frequency.

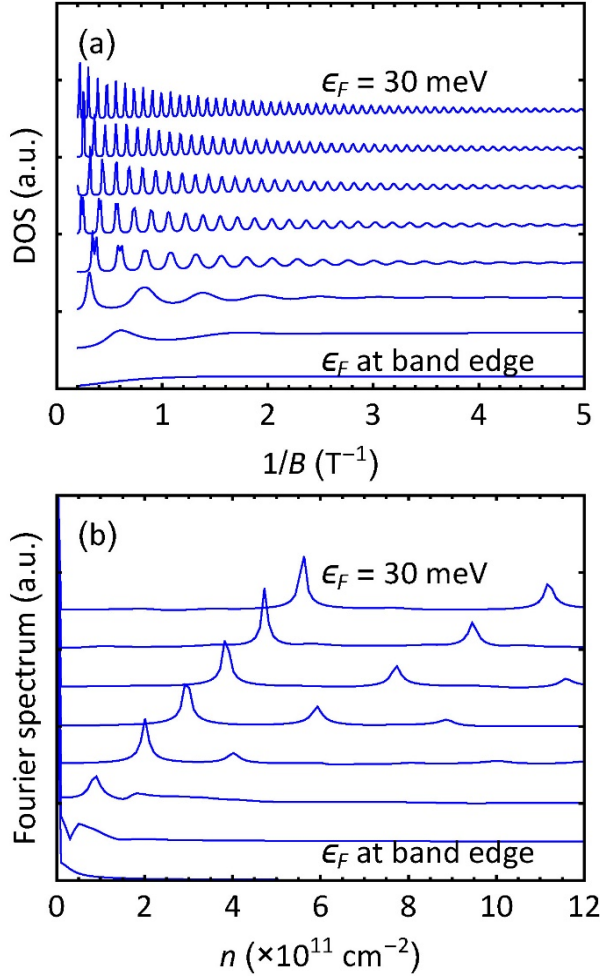

Fig. S3. (a) The DOS oscillations in the metallic regime with different Fermi energies, and (b) the Fourier spectra, where the horizontal axis is converted into the electron density $n$ with Eq. (1). Curves with different Fermi energies are offset vertically for clarity.

## B. DOS oscillations in the insulating regime, without Zeemann term

The DOS oscillations calculated in the insulating regime without including the Zeemann term are shown in Fig. S4. The self-energy is set to be $\Gamma_e = 1$ meV, and $\Gamma_h = 16$ meV. We take the second-order derivative in Fig. S4 (b) to remove the monotonic

· 21

background, and take the Fourier transformation to extract the oscillation frequency. The frequency barely changes with the Fermi energy within the hybridization gap.

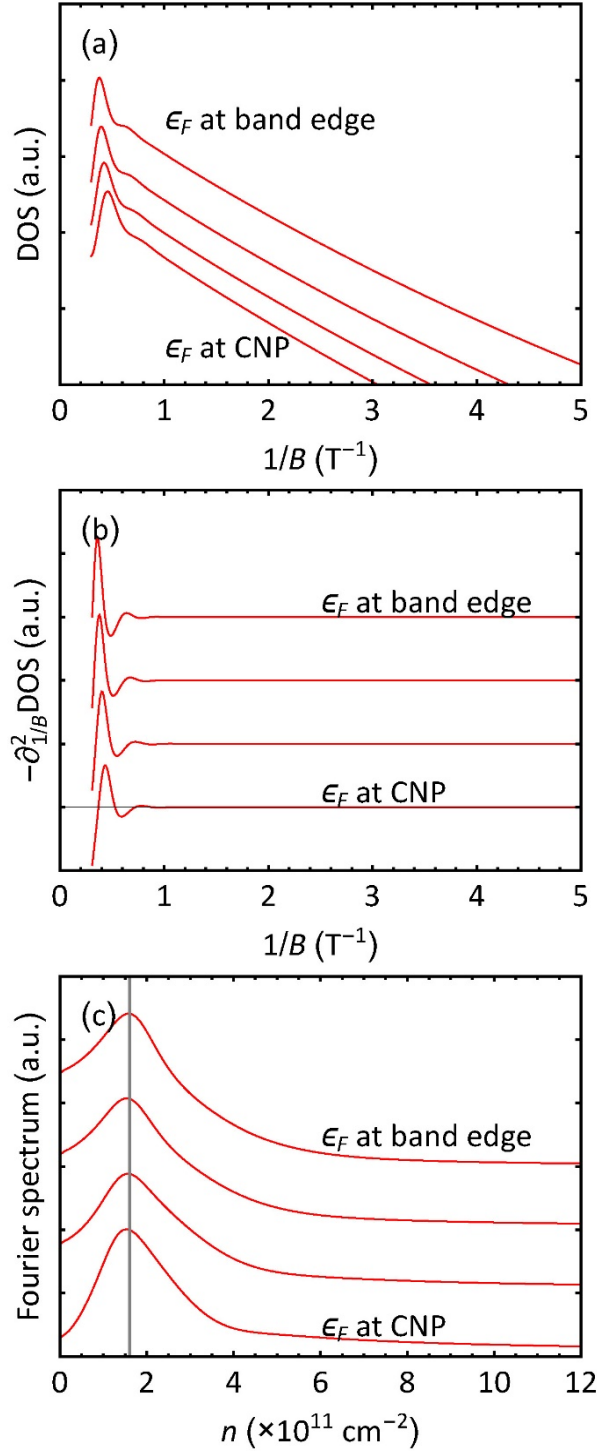

Fig. S4. (a) The DOS oscillations and (b) the second-order derivative in the insulating regime with different Fermi energies, and (c) the Fourier spectra of $-\partial^2_{B^{-1}} G_{xx}$, where the horizontal axis is converted into the electron density $n$ with Eq. (1). Curves with different Fermi energies are offset vertically for clarity. The gray line in (c) indicates the electron density at the charge neutral point (CNP) used in the calculations, $n_0 =$

$1.6 \times 10^{11}$ cm$^{-2}$.

**C. DOS oscillations in the insulating regime, with Zeemann term and inversion asymmetry**

The Zeemann splitting is important at strong magnetic fields. Moreover, a strong magnetic field also induces the level crossing of the two zeroth LLs in $|E+\rangle$ and $|H-\rangle$ sectors in the BHZ model and thus closes the energy gap. This level crossing is avoided by their hybridization, which can be induced by the bulk inversion asymmetry, the structure inversion asymmetry and particularly the perpendicular electric gating [9]. The DOS oscillation at the charge neutral point (CNP) in the presence of the Zeemann term and the hybridization of these two zeroth LLs is shown in Fig. S5. The minima in the DOS are in quantitative agreement with the conductance minima in experiments.

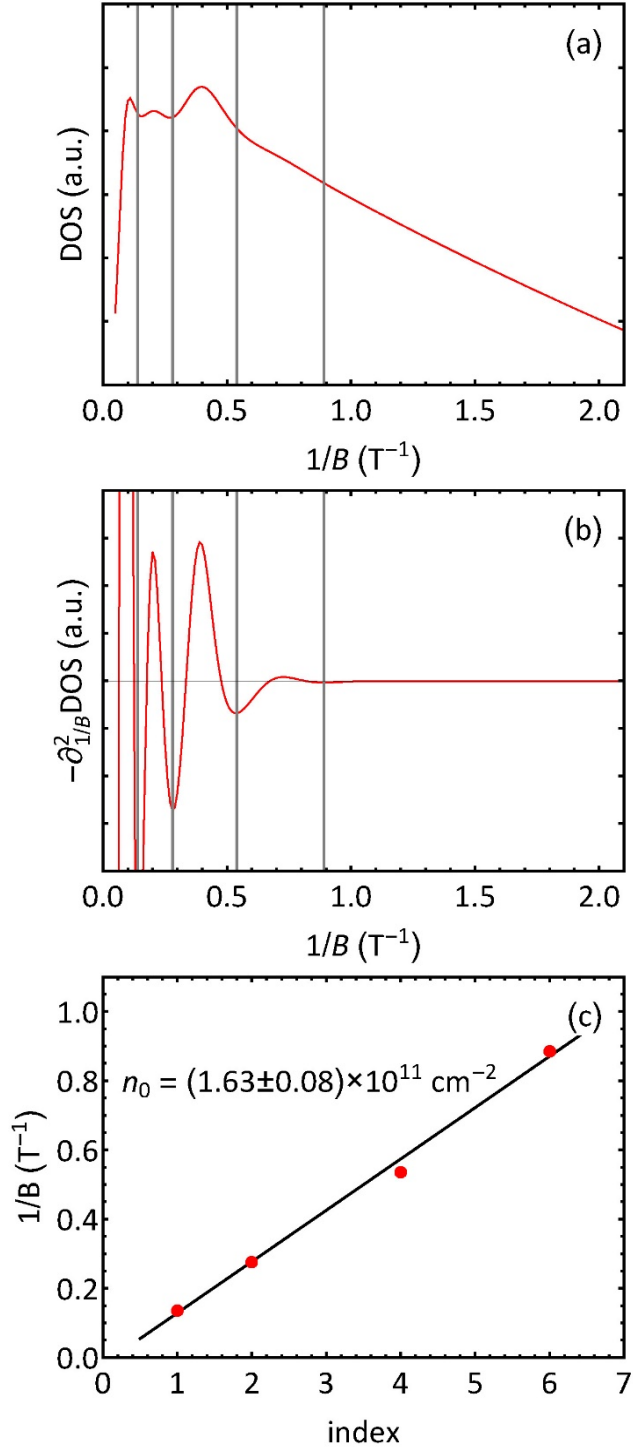

Fig. S5. (a) The DOS oscillation and (b) its second-order derivative at the CNP. Here we set $n_0 = 1.65 \times 10^{11}$ cm$^{-2}$ and the hybridization of the two zeroth Landau Levels (LLs) $\Delta = 10$ meV. (c) The LL index diagram of the DOS minima. The fitted oscillation frequency corresponds to an electron density $n_0 = (1.63 \pm 0.08) \times 10^{11}$ cm$^{-2}$.

### D. Self-energy dependence of the in-gap oscillations

From the above calculations, we can infer that the properties of the in-gap oscillations should depend on the self-energy adopted in the calculations. This is tested in our calculations and the results are shown in Fig. S6-S7.

First, as we show in Fig. S6, the in-gap DOS does not oscillate at all if the self-energy is chosen as small as in the metallic regime. This shows the necessity of a large and asymmetric self-energy in the insulating regime.

Second, we also calculate the DOS oscillations in the metallic regime assuming a large and asymmetric self-energy. The results are shown in Fig. S7. The oscillation frequency evolves smoothly into that in the insulating regime as the Fermi energy is reduced without any jump.

In summary, these results show that it is necessary to assume a large and asymmetric self-energy in the insulating regime and a small self-energy in the metallic regime in order to capture the variation of the oscillation frequency observed in the experiments.

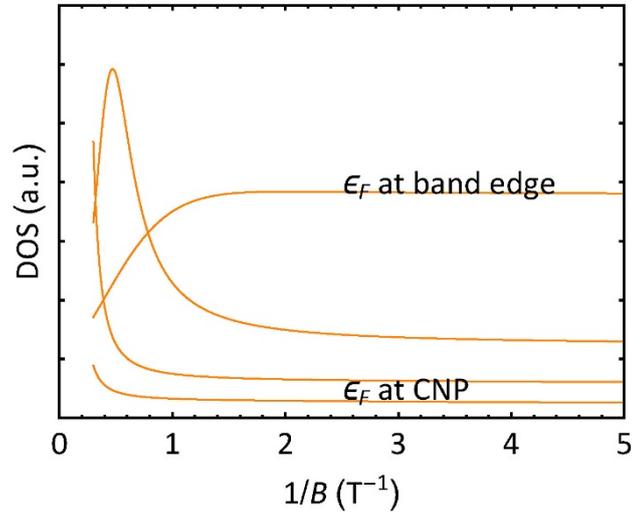

Fig. S6. The DOS in the insulating regime calculated with $\Gamma_e = 0.2$ meV and $\Gamma_h = 0.4$ meV does not show any oscillations.

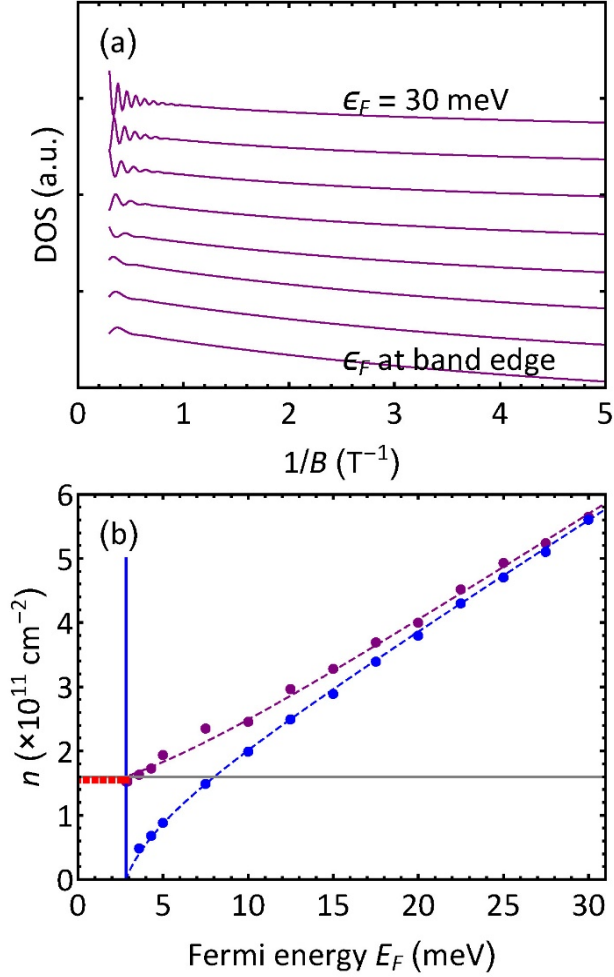

Fig. S7. (a) The DOS oscillations in the metallic regime calculated with $\Gamma_e = 1$ meV and $\Gamma_h = 16$ meV. (b) The oscillation frequency (purple dots) vs. the Fermi energy. The dashed purple line is guide for the eye. The results shown in Fig. 2d are reproduced here for comparison.

## Reference

[1] L. Du, T. Li, W. Lou, X. Wu, X. Liu, Z. Han, C. Zhang, G. Sullivan, A. Ikhlassi, K. Chang, and R.-R. Du, Phys. Rev. Lett. 119, 056803 (2017).
[2] J. Knolle and N. R. Cooper, Phys. Rev. Lett. 115, 146401 (2015).
[3] L. Zhang, X.-Y. Song, and F. Wang, Phys. Rev. Lett. 116, 046404 (2016).
[4] J. Knolle and N. R. Cooper, Phys. Rev. Lett. 118, 176801 (2017).
[5] H. Shen and L. Fu, Phys. Rev. Lett. 121, 026403 (2018).
[6] B. A. Bernevig, T. L. Hughes, and S.-C. Zhang, Science 314, 1757 (2006).
[7] C. Liu, T. L. Hughes, X.-L. Qi, K. Wang, and S.-C. Zhang, Phys. Rev. Lett. 100, 236601 (2008).
[8] X. Mu, G. Sullivan, and R.-R. Du, Appl. Phys. Lett. 108, 012101 (2016).
[9] C. Liu and S. Zhang, in *Topological Insulators*, edited by M. Franz and L. Molenkamp (Elsevier, 2013).